\documentclass[iop,twocolappendix,numberedappendix,appendixfloats]{emulateapj}
\usepackage{amsmath}
\usepackage{hyperref}
\def\teff{T_{\rm eff}} 
\shorttitle{Co-Moving Pairs}

\shortauthors{Kamdar et al.}

\begin{document}

\title{Stars that Move Together Were Born Together}

\author{Harshil Kamdar\altaffilmark{1}, Charlie Conroy\altaffilmark{1}, 
Yuan-Sen Ting\altaffilmark{2,3,4,5}, Ana Bonaca\altaffilmark{1}, Martin Smith\altaffilmark{6}, \\ Anthony G.A. Brown\altaffilmark{7}}

\altaffiltext{1}{Harvard-Smithsonian Center for Astrophysics, 
Cambridge, MA, 02138, USA} 
\altaffiltext{2}{Institute for Advanced Study, Princeton, NJ 08540, USA} 
\altaffiltext{3}{Department of Astrophysical Sciences, Princeton University, Princeton, NJ 08544, USA} 
\altaffiltext{4}{Observatories  of  the  Carnegie  Institution  of  Washington,  813  Santa Barbara Street, Pasadena, CA 91101, USA} 
\altaffiltext{5}{Hubble Fellow} 
\altaffiltext{6}{Key Laboratory for Research in Galaxies and Cosmology, Shanghai Astronomical Observatory, Chinese Academy of Sciences, 80 Nandan Road, Shanghai 200030, People’s Republic of China } 
\altaffiltext{7}{Leiden Observatory, Leiden University, P.O. Box 9513, 2300RA Leiden, The Netherlands}

\begin{abstract}

It is challenging to reliably identify stars that were born together outside of actively star-forming regions and bound stellar systems. However, co-natal stars should be present throughout the Galaxy, and their demographics can shed light on the clustered nature of star formation and the dynamical state of the disk. In previous work we presented a set of simulations of the Galactic disk that followed the clustered formation and dynamical evolution of 4 billion individual stars over the last 5 Gyr. The simulations predict that a high fraction of co-moving stars with physical and 3D velocity separation of $\Delta r < 20$ pc and $\Delta v < 1.5$ km s$^{-1}$ are co-natal. In this \textit{Letter}, we use \textit{Gaia} DR2 and LAMOST DR4 data to identify and study co-moving pairs. We find that the distribution of relative velocities and separations of pairs in the data is in good agreement with the predictions from the simulation. We identify 111 co-moving pairs in the Solar neighborhood with reliable astrometric and spectroscopic measurements. These pairs show a strong preference for having similar metallicities when compared to random field pairs. We therefore conclude that these pairs were very likely born together. The simulations predict that co-natal pairs originate preferentially from high-mass and relatively young ($< 1$ Gyr) star clusters. \textit{Gaia} will eventually deliver well-determined metallicities for the brightest stars, enabling the identification of thousands of co-natal pairs due to disrupting star clusters in the solar neighborhood.

\end{abstract}

\keywords{Galaxy: evolution -- Galaxy: kinematics and dynamics -- open clusters and associations: general}
\section{Introduction}

Close stars moving together in the Galaxy may hold valuable clues related to star formation \citep[e.g.,][]{reipurth2012formation, parker2011evolution} and the dynamical history of the Galaxy \citep[e.g.,][]{weinberg1987dynamical, monroy2014end}. Pairs of stars that are close together ($\lesssim1$ pc) are likely gravitationally bound \citep{jiang2010evolution}. Pairs further apart ($\gtrsim 1$ pc) are less likely to be gravitationally bound and may be associated with dissolving clusters \citep{kouwenhoven2010formation}, thereby allowing us to study cluster disruption and the star formation history of the Milky Way \citep[e.g.,][]{bland2010long}. 

The study of co-moving pairs using proper motions has a rich history \citep[e.g.,][]{poveda1994statistical, chaname2004disk, shaya2010very, tokovinin2012wide, alonso2015reaching}. Recent work \citep[e.g.,][]{oh2017comoving, andrews2017wide2, andrews2017wide, oelkers2017gaia, price2017spectroscopic, 2018MNRAS.480.4884E, simpson2018galah, bochanski2018fundamental, el2018wide, 2019AJ....157...78J, andrews2019using} has shown the power of using \textit{Gaia} to study co-moving pairs. The availability of 6D phase space information for millions of stars from \textit{Gaia} DR2 and complementary data from ground-based spectroscopic surveys provide a unique opportunity to understand the nature of co-moving pairs. However, {\it ab initio} simulations of galaxy formation currently do not offer the resolution necessary to interpret the small-scale phase space structure observed in {\it Gaia} data.

In \citet{kamdar2019dynamical} (hereafter K19) we presented simulations that resolve the dynamical evolution of individual stars comprising the disk over the past 5 Gyr. A key prediction of the fiducial simulation presented in K19 is the high fraction of pairs at large separations (up to $20$ pc) and at low relative velocities (up to $1.5$ km s$^{-1}$) that were born together (co-natal). In this {\it Letter} we use \textit{Gaia} DR2 and LAMOST DR4 to test the predictions presented in K19 and explore the nature of co-moving pairs. 


\begin{figure*}
 \includegraphics[width=168mm]{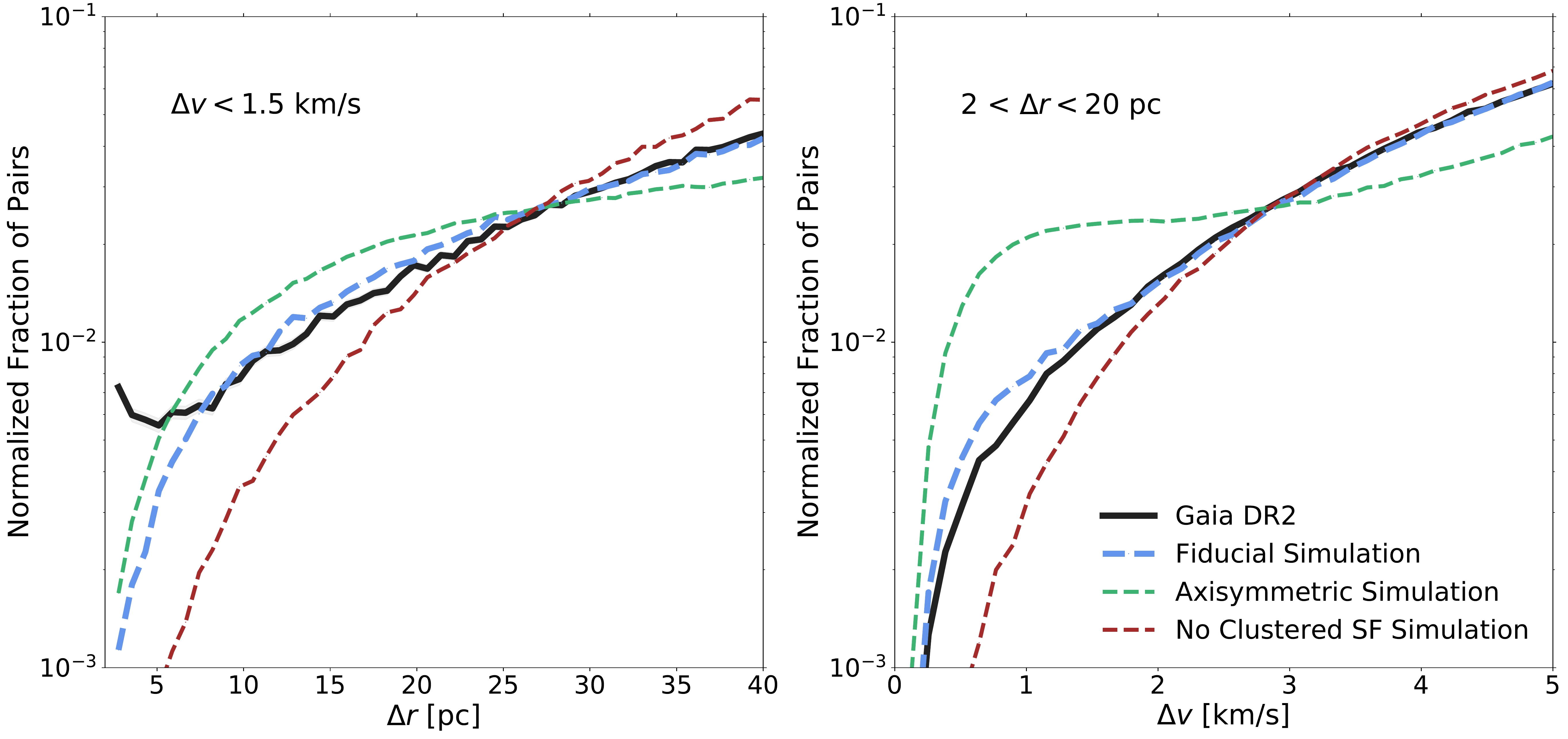}
 \caption{Normalized fraction of co-moving pairs as a function of the separation (left panel) and velocity difference (right panel) -- data is the black line, Poisson errors for the data in grey (too small to be seen), fiducial simulation is the dashed blue line, axisymmetric simulation is the dashed green line, and NCSF is the dashed red line. All three simulations underestimate clustering at the smallest separations ($\Delta r < 5$ pc). The disagreement may arise from contamination from disrupting wide binaries, or due to lack of complexity in our prescription for star formation. The differences in the fraction of pairs for the velocity difference in the simulations and the data indicates that the count of pairs at small velocities is sensitive to both clustered star formation and the non-axisymmetries in the potential of the Milky Way. The fiducial simulation, which has both, agrees better with the data than the other models. }
 \label{fig:f1}
\end{figure*}

\section{Simulations \& Data}
\label{sec:data}

In K19 we introduced a new set of simulations that were the first of their kind to model the full population of stars (younger than 5 Gyr) that comprise a Milky Way-like disk galaxy.  All stars are born in clusters with a range of initial conditions informed by observations and detailed simulations.  The dynamical evolution of 4 billion stars was performed with orbit integration of test particles coupled to a realistic time-varying galactic potential, which includes a disk, halo, bulge, bar, spiral arms, and GMCs (as perturbers). These simulations predict a rich structure in the combined phase and chemical space that should inform our understanding of the nature of clustered star formation.

The fiducial simulation presented in K19 has both non-axisymmetries in the potential (bar, spiral arms, and giant molecular clouds) and clustered star formation. We also ran two control simulations: 1) A simulation with a static axisymmetric potential and with clustered star formation. 2) A simulation with non-axisymmetric perturbations (with bar and spiral arms) but with no clustered star formation (NCSF simulation hereafter).  The three different simulations allow us to study the scales at which structure due to clustered star formation and structure due to resonances by the non-axisymmetries in the Milky Way will manifest itself on the combined chemodynamical space. 

To enable a fair comparison to \textit{Gaia} and spectroscopic surveys, we create mock catalogs of {\it Gaia} DR2-like solar neighborhoods. The solar neighborhood sphere is centered at $(-8.2, 0.0, 0.027)$ kpc and has a radius of $1$ kpc. We use the MIST stellar evolutionary tracks \citep{choi2016mesa} and the C3K stellar library (Conroy et al., in prep) to derive photometry for the simulated stars using a Kroupa IMF. \textit{Gaia} DR2 contains the radial velocities for about 7 million stars with $G_\mathrm{RVS}^\mathrm{ext}$ -- to mimic this selection, we calculate $G_\mathrm{RVS}^\mathrm{ext}$ using the Tycho color transformations presented in \citet{sartoretti2018gaia} and apply the same cut. 

An accurate error model is essential for comparisons between simulations and observations. The error in the parallax is approximated by a simple error floor of  $0.04$ mas  \citep{lindegren2018gaia}.  The dependence of  proper motion and radial velocity errors is a complex function of several parameters; we fit a gaussian mixture model (GMM) with 20 components to the combined ($G, G_{\rm{BP}}-G_{\rm{RP}}, \sigma_{\rm{\mu_{\alpha^{*}}}}, \sigma_{\rm{\mu_{\delta}}}$) and ($G, G_{\rm{BP}}-G_{\rm{RP}}, \sigma_{\rm{RV}}$) spaces respectively and sample from the conditional distributions for the respective errors given $G$ and $G_{\rm{BP}}-G_{\rm{RP}}$.

\begin{figure*}
 \includegraphics[width=168mm]{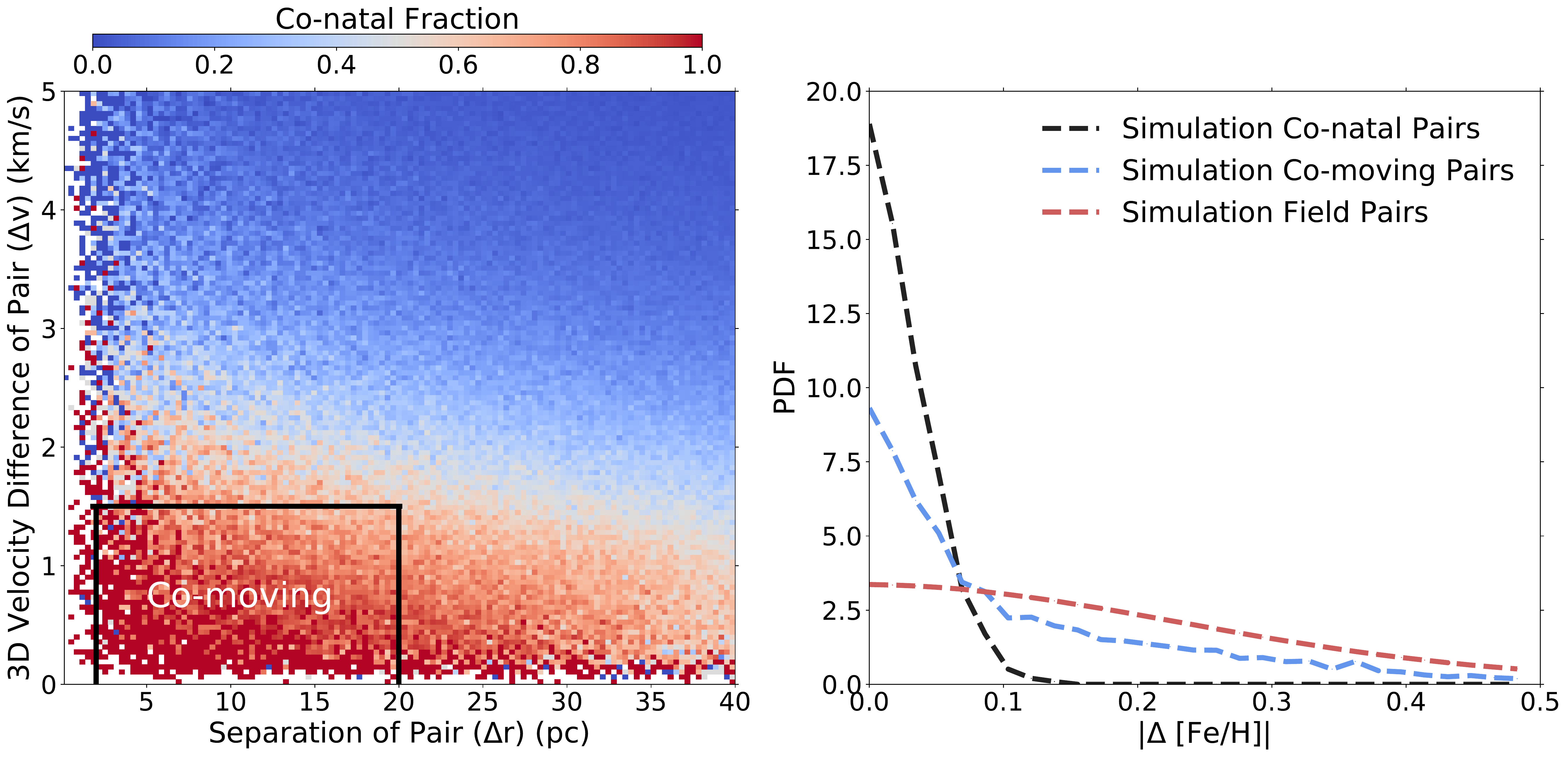} 
 \caption{Left Panel: Co-natal fraction of pairs as a function of separation ($\Delta r$) and 3D velocity difference ($\Delta v$). Pairs with $2 < \Delta r < 20$ pc and $\Delta v < 1.5$ km s$^{-1}$ have a notably high co-natal fraction -- the selection box is used to define ``co-moving" pairs. Right Panel: Expected difference in metallicites of stars in a pair.  The blue line shows the metallicity difference distribution for all co-moving pairs, while the red line shows the metallicity difference distribution for random field pairs in the solar neighborhood sample.   The black line shows the metallicity difference for the subset of co-moving pairs that are also co-natal. The simulation adopts  $\sigma_{\rm{[Fe/H]}} = 0.03$ dex.}
 \label{fig:f2}
\end{figure*}

We start with the 6D \textit{Gaia} DR2 \citep{2018A&A...616A...1G} catalog from \citet{2018MNRAS.tmp.2466M}. We cross-matched with LAMOST DR4 catalog (with duplicates removed) for the same 1 kpc sphere around the solar position as in the simulations. LAMOST DR4 \citep{deng2012lamost} provides $\teff$, [Fe/H], radial velocities, and $\log g$ using LAMOST's spectroscopic pipeline \citep{wu2011automatic, luo2015first}. Only stars that have radial velocity measurements in \textit{Gaia} DR2 are considered for the work presented here since LAMOST RVs have errors between $5-7$ km s$^{-1}$. 

We impose the following selection criteria on the \textit{Gaia} data considered in this analysis: (1) number of visibility periods $\geq 8$, (2) number of RV transits $\geq 3$, (3) re-normalized unit weight error $\leq 1.6$ (the results presented in this work do not change when making the more conservative $\leq 1.4$ cut used in \citet{LL:LL-124}), and (4) bad RVs found in \citep{2019arXiv190110460B} have been removed. We also require SNR$_i > 40$ in the LAMOST data to ensure small measurement uncertainties. After making these quality cuts, the quoted mean and median uncertainty on [Fe/H]  is 0.037 and 0.024 dex respectively -- we emphasize that it is the relative difference between metallicities that is important for this work, rather than the absolute metallicity scale. We have also identified a significant correlation between the metallicity difference and $\teff$ difference for pairs of stars, which we interpret as a systematic uncertainty on the derived metallicities. To limit this systematic uncertainty, we restrict our analysis to pairs with a temperature difference of $\Delta \teff<200$ K. 

The overall selection function for \textit{Gaia} is not critical for this work because we only consider fractional quantities when comparing between data and simulations.   We have also explored the impact of the LAMOST selection function on the population of pairs.  We find for example no statistically significant difference in the distribution of physical separations for pairs with and without LAMOST data, leading us to conclude that the LAMOST selection function does not result in a biased population of pairs.

Since the simulations in K19 do not include binarity, it is important to discuss how to isolate the phase space signature of disrupting star clusters and avoid contamination from wide binaries in the data. For our analysis, we will only consider pairs with separations greater than $2$ pc, since the Jacobi radius is $\approx 2$ pc for $\sim 1 M_{\odot}$ stars in the solar neighborhood \citep{yoo2004end, jiang2010evolution}. In addition, \citet{jiang2010evolution} predict a trough in the distribution of wide binaries between $\sim 1-10$ times the Jacobi radii, which corresponds to separations between $\sim 2-18$ pc in the solar neighborhood. Consequently, we expect the contamination from wide binaries for the spatial scales being considered in this work to be low.

Moreover, co-moving pairs could be a part of known bound clusters, moving groups or OB associations in the Milky Way. We choose to exclude pairs from these to isolate the signal of star clusters that are disrupting. The exploration of the phase space signature in data and simulations for bound objects is left for future work. Similar to \citet{oh2017comoving}, we form an undirected graph where stars are nodes, and edges between the nodes exist for co-moving pairs of stars. Consequently, a star could have multiple co-moving neighbors, and a pair of stars could be directly or indirectly connected via a sequence of edges. The graph is then split into connected components -- a connected component is a subgraph of the original graph in which any two nodes are connected to each other by a path -- to pick out only mutually exclusive pairs. This additional selection criteria efficiently filters out known open clusters, OB associations, and moving groups. 
\\[2ex]

\begin{figure*}
 \includegraphics[width=168mm]{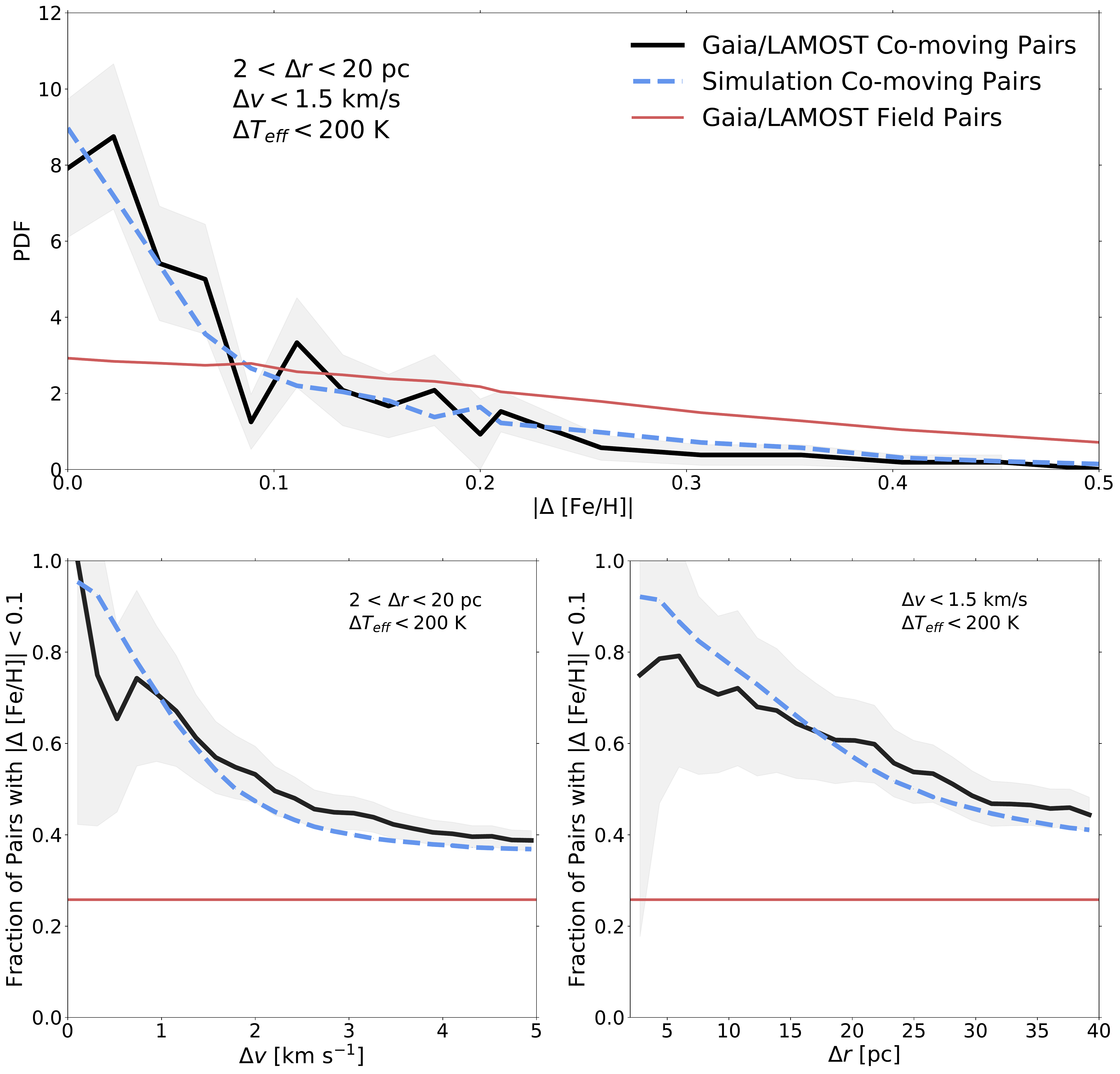}
 \caption{Top Panel: $|\Delta $[Fe/H$|$ distribution for \textit{Gaia}/LAMOST pairs with the selection cuts: $2 < \Delta r < 20$ pc and $\Delta v < 1.5$ km s$^{-1}$ (black line; with Poisson errors in grey), the simulation with the same $\Delta r$ and $\Delta v$ cuts (dashed blue line), and random field \textit{Gaia}/LAMOST pairs (red line). The metallicity difference distribution for the simulation (with $\sigma_{\rm{[Fe/H]}} = 0.03$ dex) and the data agree remarkably well. Bottom Left Panel: The fraction of pairs with $|\Delta $[Fe/H$| < 0.1$ dex for data (black line; Poisson errors in grey) and the fiducial simulation (blue dashed line) as a function of different velocity difference cuts after already imposing $\Delta r < 20$ pc. Both the simulation and the data show a similar peak of low metallicity difference pairs at low velocity differences. As $\Delta v$ increases, the fraction of pairs with $|\Delta$ [Fe/H$| < 0.1$ approaches the value for random field pairs (red line). Bottom Right Panel: The fraction of pairs with $|\Delta $[Fe/H$| < 0.1$ dex for data (black line; Poisson errors in grey) and the fiducial simulation (blue dashed line) as a function of different separation cuts after already imposing $\Delta v < 1.5$ km s$^{-1}$. Both the simulation and the data show a similar peak of low metallicity difference pairs at low separations. As $\Delta r$ increases, the fraction of pairs with $|\Delta $[Fe/H$| < 0.1$ approaches the value for random field pairs (red line).}
 \label{fig:f3}
\end{figure*}

\section{Co-moving Pairs in Simulations \& Data}
\label{sec:results}

In this section we investigate co-moving pairs in the simulations and the data. We begin with a pair-wise comparison of the fraction of pairs as a function of separation and velocity difference in the simulations and in the data in Figure \ref{fig:f1}. The left panel of Figure \ref{fig:f1} shows the fraction of co-moving pairs as a function of $\Delta r$ for the different simulations (dashed lines of different colors) and the data (black solid line) after imposing a $\Delta v < 1.5$ km s$^{-1}$ cut. At smaller separations, clustering seems to be stronger in the data than in the simulations. The discrepancy at the smallest separations could be due to contamination from wide binaries or hint at complexities in the data regarding star formation that are not adequately modeled in the fiducial simulation. 

The right panel of Figure \ref{fig:f1} shows the fraction of co-moving pairs as a function of $\Delta v$ for the different simulations and the data with the cut $2 < \Delta r < 20$ pc. The axisymmetric simulation has a notably higher fraction of pairs at lower $\Delta v$ due to the absence of any large-scale scattering. The NCSF simulation has a notably lower relative fraction at lower $\Delta v$ due to the absence of clustered star formation. The fiducial simulation, which has both clustered star formation and non-axisymmetries in the potential, is in excellent agreement with the data. 

Figure \ref{fig:f2} shows the co-natal fraction of co-moving pairs in the fiducial simulation presented in K19. The simulation predicts that pairs of stars with a velocity difference up to $1.5$ km s$^{-1}$ but a high separation of up to $20$ pc are highly likely to be born together. The large physical separation suggests that the co-moving pairs of stars are likely part of a disrupting star cluster -- studying such co-natal pairs in data will provide key insights into star formation in the disk. Consequently, using the simulation as a prior of sorts, we will use the selection box $2 < \Delta r < 20$ pc and $\Delta v < 1.5$ km s$^{-1}$ to both avoid most wide binaries and look for signatures of dissolving star clusters in the data.   

While, as we can see in the left panel of Figure \ref{fig:f2}, the selection box contains a large number of co-moving pairs that are co-natal, there is still some contamination (about $\sim$20\%) from field pairs. To test whether we find a similarly high co-natal fraction in data, we can use metallicities since stars born in the same cluster are believed to have essentially identical metallicities \citep[e.g.,][]{de2007chemical, bovy2016chemical}, modulo atomic diffusion \citep{dotter2017influence}. The right panel of Figure \ref{fig:f2} shows expectations for the metallicity ([Fe/H]) difference distribution of co-moving pairs. The black line shows pairs of stars known to be co-natal in the selection box, the blue line shows all pairs of stars in the selection box, and the red line shows random field pairs. The measurement uncertainty in [Fe/H] in the simulations is $0.03$ dex. The selection criteria are effective at picking out co-natal pairs but there is still enough contamination from field pairs for there to be a tail at larger metallicity differences. The blue line will be used to compare to the \textit{Gaia}/LAMOST cross-match. 

\begin{figure*}
 \includegraphics[width=168mm]{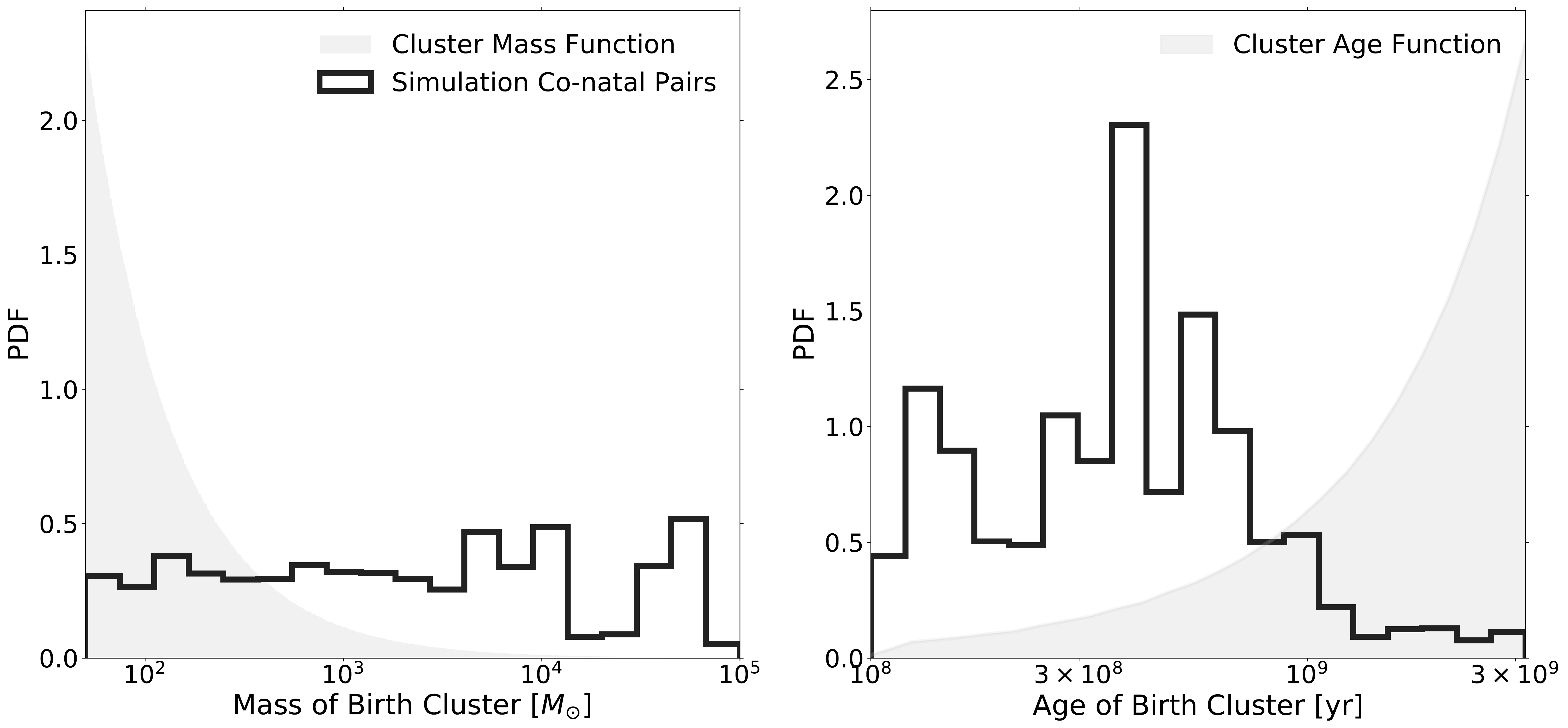}
 \caption{Demographics of simulated co-moving pairs that are co-natal. Left Panel: The distribution of the masses of the birth clusters of the co-natal pairs in the simulation (black) compared to the overall cluster mass function (grey). The birth cluster masses for co-natal pairs peaks at significantly higher masses -- this is due to the many more pairs of stars that are possible as the cluster mass increases. Right Panel: The distribution of ages of the co-natal pairs in simulation (black) compared to the overall cluster age function (grey).}
 \label{fig:f4}
\end{figure*}

The top panel of Figure \ref{fig:f3} shows the metallicity difference ($|\Delta$ [Fe/H]$|$) for co-moving pairs with $2 < \Delta r < 20$ pc and $\Delta v < 1.5$ km s$^{-1}$. The agreement between the data and the simulation (which has a fiducial $\sigma_{\rm [Fe/H]} = 0.03$ dex as a mock observational uncertainty) is very good. Moreover, the distribution of metallicity differences is significantly narrower than the field metallicity difference distribution, suggesting that the co-moving pairs we have identified are co-natal. 

The bottom left panel of Figure \ref{fig:f3} shows the fraction of pairs with $|\Delta$ [Fe/H]$| < 0.1$ as a function of different velocity cuts after already imposing $2 < \Delta r < 20$ pc. The fraction of stars with a low metallicity difference quickly drops for velocity differences larger than $\sim 1-1.5$ km s$^{-1}$ in both the data and the simulation. The fraction of pairs with low metallicity differences approaches that of the field pairs' at progressively larger $\Delta v$ cuts. 

The bottom right panel of Figure \ref{fig:f3} shows the fraction of pairs with $|\Delta$ [Fe/H]$| < 0.1$ as a function of different separation cuts after already imposing $\Delta v < 1.5$ km s$^{-1}$. The fraction of stars with a low metallicity difference drops relatively smoothly as a function of separation for both the simulation and the data -- this trend is also clearly visible in Figure \ref{fig:f2}. The fraction of pairs with low metallicity differences approaches that of the field pairs' at progressively larger $\Delta r$ cuts.  Taken together, the results in Figure \ref{fig:f3} indicate that the final selection cuts of $2 < \Delta r < 20$ pc and $\Delta v < 1.5$ km s$^{-1}$, lead to a fairly clean sample ($\sim 80\%$ pure) of co-moving pairs that were born together. 

The simulation offers the opportunity to identify properties of the birth sites of the co-natal pairs.  The left panel of Figure \ref{fig:f4} shows the mass of the birth cluster for the co-natal co-moving pairs in the selection box in Figure \ref{fig:f2} compared to the overall cluster mass function (CMF). The clusters that give birth to co-natal co-moving pairs have notably higher masses relative to the overall CMF. This is due to the much larger number of pairs of stars that are possible with an initially large number of stars in the cluster.  The right panel of Figure \ref{fig:f4} shows the age of the birth cluster for the co-natal co-moving pairs in the selection box shown in Figure \ref{fig:f2} compared to the overall cluster age function. The clusters that give birth to co-natal co-moving pairs are mostly younger than $1$ Gyr. The age distribution of these clusters offers a noteworthy parallel to the discussion of visibility timescale presented in K19, where we argued that the phase space signature of stars born in clusters should be visible up to $1$ Gyr.

\begin{deluxetable*}{rrrrr} 
\tabletypesize{\footnotesize} 
\tablecolumns{5} 
\tablewidth{0pt} 
\tablecaption{Catalog of co-moving pairs analyzed in this work. The catalog is available in its entirety for download \href{http://harshilkamdar.github.io/2019/04/03/pairs.html}{here}.} 
\tablehead{ 
Star 1 \textit{Gaia} DR2 ID & Star 2 \textit{Gaia} DR2 ID & $\Delta r$ (pc) & $\Delta v$ (km s$^{-1}$) & $|\Delta$ [Fe/H]$|$ (dex) \\
\vspace{-0.2cm} }
\startdata 

454335594920988288 & 454131631217853312 & 19.518 & 0.363 & 0.041 \\
2130506398193844352 & 2128941861868381952 & 18.991 & 1.347 & 0.043 \\
1897440689367972096 & 1897440689367972736 & 4.447 & 1.427 & 0.060 \\
946471222781159040 & 946435007617437056 & 6.852 & 0.978 & 0.630 \\
893835131555196928 & 881823826013408384 & 7.434 & 0.901 & 0.059 \\
3992114209767880960 & 3994508877374172160 & 17.760 & 1.249 & 0.007 \\

\enddata 
\vspace{-0.3cm} 
\label{table:catalog}
\end{deluxetable*}

\section{Summary}
\label{sec:summary}

In this {\it Letter} we have presented evidence that co-moving pairs of stars (identified as having relative physical separation $2 < \Delta r < 20$ pc, and relative velocity $\Delta v < 1.5$ km s$^{-1}$)  were very likely born together. Moreover, we have shown that the distribution of pair-wise velocities and physical separation is sensitive to both clustered star formation and non-axisymmetries of the Galactic potential. The simulation presented in K19 is able to reproduce both of these distributions. Furthermore, the simulation predicts that the co-natal co-moving pairs were born in preferentially high-mass and relatively young clusters relative to the field population. Table \ref{table:catalog} lists all 111 pairs in our final catalog and their properties. 

The primary metric used here for determining whether a pair is co-natal is the metallicity difference ($|\Delta $[Fe/H]$|$). For this we used data from the low resolution LAMOST spectroscopic survey since it contains far more stars in the solar neighborhood than other surveys. Follow-up high resolution spectra of these co-moving pairs is required to confirm the chemical similarity of the pairs, which in turn would bolster the argument that the pairs share a common birth site. \textit{Gaia} DR4 will deliver SNR $\sim 50$ spectra for stars with $G \sim 12$, allowing calculations of [Fe/H] with uncertainties $\leq 0.05$ dex \citep{2016A&A...585A..93R, ting2017measuring}. The simulations predict that we could study thousands of co-moving pairs due to disrupting star clusters in the solar neighborhood using \textit{Gaia} alone. 

Co-moving pairs offer novel constraints on the nature of clustered star formation and the recent dynamical history of the disk. However, pairs offer a limited ($N=2$) view of the phase space structure of stars in the disk. Additional insight will be gained by considering the general clustering properties of stars in various phase space projections.  This will be the subject of future work.

\acknowledgements
We thank Adrian Price-Whelan, Kareem El-Badry, Yan-Fei Jiang, Semyeong Oh, and Hans-Walter Rix for useful discussions. The computations in this paper were run on the Odyssey cluster supported by the FAS Division of Science, Research Computing Group at Harvard University. HMK  acknowledges  support  from  the  DOE  CSGF  under  grant  number DE-FG02-97ER25308. CC acknowledges support from the Packard Foundation. YST  is supported  by the NASA Hubble Fellowship grant HST-HF2-51425.001 awarded by the Space Telescope Science Institute. MCS acknowledges financial support from the National Key Basic Research and Development Program of China (No. 2018YFA0404501) and NSFC grant 11673083. 

This work has made use of data from the European Space Agency mission \textit{Gaia}, processed by the \textit{Gaia} Data Processing and Analysis Consortium (DPAC). Funding for the DPAC has been provided by national institutions, in particular the institutions participating in the \textit{Gaia} Multilateral Agreement. Guoshoujing Telescope (the Large Sky Area Multi-Object Fiber Spectroscopic Telescope LAMOST) is a National Major Scientific Project built by the Chinese Academy of Sciences. Funding for the project has been provided by the National Development and Reform Commission. LAMOST is operated and managed by the National Astronomical Observatories, Chinese Academy of Sciences.


\bibliographystyle{yahapj}

\bibliography{main}

\end{document}